\documentclass[showkeys,preprintnumbers,amsmath,amssymb]{revtex4}
\usepackage{graphicx}
\usepackage{graphics}
\usepackage{epsfig}

 \def\be{\begin{equation}}
 \def\ee{\end{equation}}
 \def\ba{\begin{array}}
 \def\ea{\end{array}}

 \newcommand{\intt}{{{\int_{0}}^{\infty}}}
 \newcommand{\inta}{{{\int_{0}}^{a}}}

\begin{document}

\title[Elastic contact to a coated half-space - Effective elastic modulus and real penetration]
 {Elastic contact to a coated half-space - Effective elastic modulus and real penetration}

\author{ A. Perriot, E. Barthel }

\address
{Laboratoire CNRS/Saint-Gobain "Surface du Verre et Interfaces",
39, quai Lucien Lefranc, BP 135, F-93303 Aubervilliers Cedex,
France}

\email{antoine.perriot@saint-gobain.com,
etienne.barthel@saint-gobain.com}


\date{\today}

%%% ----------------------------------------------------------------------

\begin{abstract}
A new approach to the contact to coated elastic materials is
presented. A relatively simple numerical algorithm based on an
exact integral formulation of the elastic contact of an
axisymmetric indenter to a coated substrate is detailed. It
provides contact force and penetration as a function of the
contact radius. Computations were carried out for substrate to
layer moduli ratios ranging from $10^{-2}$ to $10^{2}$ and various
indenter shapes. Computed equivalent moduli showed good agreement
with the Gao model for mismatch ratios ranging from 0.5 to 2.
Beyond this range, substantial effects of inhomogeneous strain
distribution are evidenced. An empirical function is proposed to
fit the equivalent modulus. More importantly, if the indenter is
not flat-ended, the simple relation between contact radius and
penetration valid for homogeneous substrates breaks down. If
neglected, this phenomenon leads to significant errors in the
evaluation of the contact radius in depth-sensing indentation on
coated substrates with large elastic modulus mismatch.
\end{abstract}

%%% ----------------------------------------------------------------------
\maketitle
%%% ----------------------------------------------------------------------

\section{Introduction}

With the increasing use of coated and multilayered systems, there
is a growing need for accurate measurement of the mechanical
properties of thin films. Amongst the few possible techniques,
nanoindentation appears as the most promising. However, the
methods used to analyze indentation data (the most commonly used
being that by Oliver and Pharr$^1$) are based on models valid for
isotropic homogeneous elastic solids$^{2,3}$. As a consequence,
the so-called ``equivalent elastic modulus'', $E_{\rm eq}$,
obtained through these methods is a combination of the respective
moduli of the film, $E_{\rm film}$, and of the substrate, $E_{\rm
substrate}$. The relative weight of each material, as was pointed
out by Doerner and Nix$^4$, varies with the penetration depth.

Several models have been proposed to extract intrinsic material
properties of the film from this depth-dependent equivalent
modulus. Most of them are empirical models based on the following
structure: \be E_{\rm eq}=E_{\rm substrate}+(E_{\rm film}-E_{\rm
substrate})\,\Phi(x) \ee in which $x$ is the ratio of the contact
radius, $a$, or the contact depth, $h_{c}$, to the film thickness,
$t$, and $\Phi$ is the ``weight function" which equals 1 when $x$
is 0 and 0 when $x$ is infinite. The most commonly-used of these
models have been summarized and studied experimentally by
\nolinebreak Men$\breve{c}$ik  et al.$^5$.

Analytical approaches have also been proposed. The method
introduced by Gao et al.$^6$ is based on a perturbative
calculation of the elastic energy of a coated substrate indented
with a flat punch. This approach is based on the assumption that
the mechanical properties of both materials do not differ widely.
Comparisons with finite-element (FE) calculations proved that this
model provides a precise account for load/displacement responses,
at least for modulus ratios up to 2. Another approximate approach
by Yoffe$^7$ focuses on the calculation of an additional stress
field which would compensate the inhomogeneity of the coated
solid during its indentation by a rigid flat punch. From the
latter, an approximate formula linking the global compliance to
the geometry of the system is deduced. A third approach, allowing
the analytically exact calculation of the stress field in the
whole coated material, has been proposed by Schwarzer$^8$. Based
on results by Fabrikant$^9$, it uses an electrostatic-like method
of images approach, leading to the calculation of the sum of an
infinite series.

In the present work, we propose an alternative method relying on a
previous work by Li and Chou$^{10}$, in which they calculated the
Green function for a coated substrate. Unfortunately, their
stress/strain relation could not be inverted, thus proving of
little use for contact problems. Using the auxiliary fields
introduced by Sneddon$^{3,11}$ and later on developed by Huguet
and Barthel$^{12}$ and Haiat and Barthel$^{13}$, we first
reformulated their expression to allow the problem to be inverted
at low numerical cost.This approach is developed in Part I.

In Part II, we present the results of our calculations. We first
compare our results for the equivalent modulus with the Gao model.
Following that, we propose a new fitting function for the contact
radius dependence of the equivalent modulus.

As a result for sphere and cone indentation, a non-trivial
penetration/contact radius relation is obtained, which differs
markedly from the Hertz equation. The implications of these
results for data treatment in depth-sensing indentation methods to
determine the modulus of very thin films are discussed.

\part{Algorithm}

\section{Green function of a coated substrate}

Let us consider a mechanical system composed of a coated elastic
half-space under an axisymmetric frictionless loading (Fig. 1).
The layer, of thickness $t$, and the half-space are supposed
elastic, isotropic and homogeneous, while their adhesion is
supposed to be perfect. Let $E_{0}$ and $\nu_{0}$ (resp. $E_{1}$
and $\nu_{1}$) be the elastic modulus and the Poisson ratio of the
half-space (resp. of the layer).

Using Hankel transforms, in the framework of linear elasticity, Li
and Chou$^{10}$ obtained a relation between the applied normal
stress $q(r)$ at the surface of the layer (taken positive when
compressive) and the normal displacement $u(r)$ (being positive
when inwards) in the whole system. Particularizing this general
relation between $q$ and $u$ for the surface plane (i.e. for
$z=0$), they obtained the following expression:
\begin{equation}\label{equil}
u(r)={{\int_{0}}^{\infty}}dk\,{\overline{q}}(k)J_{0}(kr)\mathcal{C}(kt)
\end{equation}
in which we have :
$$\mathcal{C}
(kt)={\frac{2}{{E_1^*}}}{\frac{1+4b\,kt\,{e}^{-2kt}-ab\,{e}^{-4kt}}{1-(a+b+4b{(kt)}^{2}){e}^{-2kt}+ab\,{e}^{-4kt}}}$$
\begin{center}
$a={\frac{\alpha\gamma_{3}-\gamma_{1}}{1+\alpha\gamma_{3}}}$,
   $b={\frac{\alpha-1}{\alpha+\gamma_{1}}}$,
   $\alpha={\frac{E_1(1+\nu_0)}{E_0(1+\nu_1)}}$,
   $\gamma_{1}=3-4{\nu}_{1}$ and $\gamma_{3}=3-4{\nu}_{0}$
\end{center}
${E_1^*}$ being the reduced modulus of the layer defined as
$E_1/(1-{\nu_1}^2)$, $J_{0}(x)$ the 0$^{th}$-order Bessel function
of the first kind and ${\overline{q}}$ the 0$^{th}$-order Hankel
transform of $q$ defined as\nolinebreak :
$${\overline{q}}(k)=\intt dr\,r\,J_{0}(kr)\, q(r)$$

Knowing the loading on the whole surface of the system, one may
calculate exactly the complete surface displacement.
Unfortunately, indentation problems are characterized by mixed
boundary conditions : one only knows the surface displacement
under the contact and the applied stress outside of it. Thus, Eq.
(\ref{equil}), though analytically exact, is of little use in this
form. In the following, we propose a reformulation of this
expression, which, using auxiliary fields, allows us to bypass
this difficulty.

\section{Introducing auxiliary fields}

In order to present a model for the adhesive contact of
viscoelastic spheres, Barthel and Haiat$^{13}$ introduced the
auxiliary fields $g$ and $\theta$ defined as the following cosine
Fourier transforms of the Hankel transforms of the normal surface
stress $q(r)$ and displacement $u(r)$ respectively :
\begin{equation}\label{g_Hank}
g(s)=\intt dk\, {\overline{q}}(k)\cos (ks)
\end{equation}
\begin{equation}\label{th_Hank}
\theta(s)=\intt dk\, k {\overline{u}}(k)\cos (ks)
\end{equation}
Let us first rewrite Eq. (\ref{equil}) with the Hankel transform.
We obtain :
\begin{equation}\label{equil_Hank}
k {\overline{u}}(k)=\mathcal{C}(kt){\overline{q}}(k)
\end{equation}
Let us now apply the cosine Fourier transform to Eq.
(\ref{equil_Hank}). After some rewriting, we get the following
expression \nolinebreak:
\begin{equation}\label{equil_couche}
\theta(s)={\frac{2}{\pi}}\intt  g(r) \left(\intt dk\, \mathcal{C}
(kt) \cos (kr) \cos (ks)\right)dr
\end{equation}
A few comments on Eq. (\ref{equil_Hank}) and Eq.
(\ref{equil_couche}) are now in order.

\section{General comments on the equations}

First, we notice that, when turning the system into an homogeneous
one, either using the same material for the layer and the
substrate or making $t$ nil or infinite, Eq. (\ref{equil_Hank})
and Eq. (\ref{equil_couche}) turn into :
\begin{equation}\label{equil_hom_Hank}
k{\overline{u}}(k)={\frac{2}{E^{*}}}{\overline{q}}(k)
\end{equation}
\begin{equation}\label{equil_hom_real}
\theta(r)={\frac{2}{E^{*}}}g(r)
\end{equation}

These are the equations given by Huguet and Barthel$^{12}$ in the
case of an homogeneous half-space. The interesting point is that
the relation between $\theta$ and $g$ is local ({\it i.e.}
diagonal), in contrast to Eq. (\ref{equil_couche}). Inversion is
therefore straightforward.

Moreover, expressing Eq. (\ref{g_Hank}) and Eq. (\ref{th_Hank}) in
the real space, we obtain :
\begin{equation}\label{g_real}
g(s)={{\int_{s}}^{\infty}}dr
{\frac{rq(r)}{\sqrt{{r}^{2}-{s}^{2}}}}
\end{equation}
\begin{equation}\label{th_real}
\theta(s)={\frac{d}{ds}}{{\int_{0}}^{s}}dr
{\frac{ru(r)}{\sqrt{{s}^{2}-{r}^{2}}}}
\end{equation}

Considering the integration limits, one notices that $g$ (resp.
$\theta$) depends on $q(r)$ (resp. $u(r)$) only for \linebreak
$r\geq s$ (resp. $r\leq s$). Thus, $g$ and $\theta$ appear to be
well-suited to contact problems.

In the case of an homogeneous material, a combination of Eq.
(\ref{equil_hom_real}), Eq. (\ref{g_real}) and Eq. (\ref{th_real})
allows to solve the problem easily.

In principle, similar auxiliary functions can be built to
diagonalize Eq. (\ref{equil_couche}) for the coated system. In
practice, as explicit expressions have not been obtained yet, we
introduced a numerical method to invert Eq. (\ref{equil_couche}).

\section{A stress/displacement relation for indentation of a coated solid by a rigid indenter}

\subsection{Problem definition}

Let us now consider Eq. (\ref{equil_couche}) for application to an
indentation experiment. Let us not make any hypothesis on the
shape of the indenter, apart from the fact that it is rigid,
convex, axisymmetric and frictionless. For simplicity, we will
consider the contact between the indenter and the coated material
to be non-adhesive.

The boundary conditions of this problem are the following :
\be\left\{\ba{rl}\label{boundary}
\displaystyle \forall r \leq a,&u(r)=\delta-h(r)\\
\displaystyle \forall r \geq a,&q(r)=0
 \ea\right.\ee
where $h(r)$ the shape of the indenter and $a$ the contact radius.

Note that this type of loading, because it only considers the
normal displacement, does not model the indenter shape exactly, as
has been shown by Hay et al.$^{14}$. The minor corrections taking
into account the radial displacement will not be considered here.

Combining Eq. (\ref{g_real}) and Eq. (\ref{boundary}), we have :
\be \forall \, r\geq a, g(r)=0 \ee

Eq. (\ref{equil_couche}) then becomes :
\begin{equation}\label{equil_contact}
\theta (s)={\frac{2}{\pi}}\inta g(r) \left(\intt dk\, \mathcal{C}
(k) \cos (kr) \cos (ks)\right)dr
\end{equation}
while $\theta(r)$ is known on $[0;a]$ through Eq. (\ref{th_real})
and Eq. (\ref{boundary}).

Then, under the contact, our indentation problem turns into an
integral equation of the type $h(r)=\inta ds f(s)K(r,s)$ where $h$
and $K$ are known.

As the method is similar whatever the indenter shape, we will only
detail here the calculation in the case of the cone indentation,
while the case of the flat punch and of the sphere will be
developed in appendix A.

\subsection{Cone indentation}
Under the contact, the given normal surface displacement is
$u(r)=\delta-r/\tan (\omega)$, where $\omega$ is the half-included
angle of the cone. We deduce that :
\begin{equation}\label{th_cone}
 \forall\, s \leq a,\, \theta(s)=\delta-{\frac{\pi s}{2 \tan (\omega)}}
\end{equation}

Let us now normalize the contact variable. The appropriate
characteristic length in the case of a homogeneous substrate is
$a$, as appears in the Hertz theory. When turning an homogeneous
substrate into a coated one, we introduce another characteristic
scale which is the layer thickness $t$. Thus, we expect an
appropriate normalization to exhibit the ratio $a/t$. Indeed, one
notes that ${ta^2}/{a^3}$ is an estimate of the fraction of the
``elastically active volume'' in the film. Normalizing the length
by $a$ also normalizes all the other quantities by the values
obtained for the homogeneous material with the mechanical
characteristics of the layer.

If we now introduce the function $\mathcal{Z}$, defined as
follows, $\mathcal{Z}(x)={\frac{E_1^*}{2}}\mathcal{C}(x)-1$, we
obtain :
\begin{equation}%10
\theta(s)={\frac{2}{E_1^*}}g(s)+{\frac{2}{\pi}}\inta  g(r)
\left(\intt dk\, {\frac{2}{E_1^*}}\mathcal{Z}(kt) \cos (kr) \cos
(ks)\right)dr
\end{equation}

This expresses the global response of the system to mechanical
stress as the reaction of a semi-infinite film to which is added a
corrective term quantifying the substrate effect. This expression
is similar to the approach of the problem by Yoffe$^7$ in the
specific case of a flat punch indentation of a coated material.

We then introduce the following normalized quantities :

\begin{equation}
\rho\equiv{\frac{r}{a}};\,\varsigma\equiv{\frac{s}{a}};\,\tau\equiv{\frac{t}{a}};\,\eta\equiv
ka
 \ee

 \be \Delta \equiv{\frac{2\delta \tan(\omega) }{\pi a}}
\ee

\be \mathcal{Z}(x)\equiv{\frac{E_1^*}{2}}\mathcal{C}(x)-1 \ee

\be G(r)\equiv{\frac{4 \tan(\omega) g(r)}{\pi a {E_1^*}}} \ee

Then Eq. (13) reads :

\begin{equation}%19
\forall\, \varsigma \leq 1,\,
\Delta-{\varsigma}=G(\varsigma)+{\frac{2}{\pi}}{{\int_{0}}^{1}}
G(\rho) \left(\intt d\eta\, \mathcal{Z} (\eta \tau) \cos(\eta
\rho) \cos (\eta \varsigma)\right)d\rho
\end{equation}
\newpage
The normalized applied force is :

\begin{equation}%20
\Pi={\frac{4 \tan(\omega)}{\pi a^2
E_1^*}}P=4{{\int_{0}}^{1}}d\rho\, G(\rho)
\end{equation}

And the normalized equivalent modulus is :

\begin{equation}%21
{\mathcal{E}}_{\rm eq}^*={\frac{E_{\rm
eq}^*}{E_1^*}}={\frac{\Pi}{2{\Delta}^2}}
\end{equation}

\section{Numerical treatment of the equations}

\subsection{Implementation of the model}

Let us consider Eq. (20) and discretize $[0;1]$ into $N$. From now
on, we shall use $\varsigma_i={{i}/{N}}$ and $\rho_j={{j}/{N}}$.
If we approximate the integral on $\rho$ by a discrete sum, we
then get :

\begin{equation}%25
\Delta-\varsigma_i=G(\varsigma_i)+{\frac{1}{N\pi}}G(0)K(\varsigma_i,0,\tau)+{\frac{1}{N\pi}}G(1)K(\varsigma_i,1,\tau)+{\frac{2}{N\pi}}\sum_{j=1..N-1}G(\rho_j)K(\varsigma_i,\rho_j,\tau)\end{equation}
with $K(\varsigma,\rho,\tau)=\intt d\eta\, \mathcal{Z}
(\eta \tau) \cos(\eta \rho) \cos (\eta \varsigma)$\\

From Eq. (12) we have $G(1)=0$. Then, for a given $\tau$ --- that
is to say for a given contact radius --- we have to solve the
$(N+1)\times(N+1)$ linear system introduced in Eq.
\nolinebreak(23) for the remaining $N$ values of the $G$ field and
the normalized penetration $\Delta$. The
$\underline{\underline{K}}$-matrix elements can be calculated with
a Fast Fourier Transform (FFT) algorithm for numerical efficiency.
Finally, the applied load $\Pi$ and equivalent modulus
$\mathcal{E^*_{\rm eq}}$ are obtained through Eq. (21) and Eq.
(22).

\subsection{Numerical considerations}

The accuracy of the numerical solution depends on the dimension
$N$ of the $\underline{\underline{K}}$ matrix (which is associated
to the discretization of $[0;1]$), the cut-off $B$ for the
sampling range of the $\mathcal{Z}$ function and the sampling rate
$B/2^k$ for the FFT calculation of the matrix elements.

As $\mathcal{Z}$ decreases in an exponential-like way and tends to
zero when $\eta$ becomes infinite, it is possible to choose a
rather small value for $B$. We first considered the case $a/t=100$
and chose $B=1000$. We then tested increasing values of $N$ and
$k$. Our results appear to converge with a deviation smaller than
1$\%$ when $N$ is greater than 700 (for a given $k$ of 19) and
when $k$ is greater than 14 (with $N$=700).

However, with $N=700$ and $B=1000$, the output returned for small
values of $a/t$ converges only for $k=20$, as the decay length of
$\mathcal{Z}$ decreases with $a/t$. This leads to an increase in
term of calculation time. Thus, we decided to fix $N=700$ and
$k=14$, which constitutes a good compromise between calculation
time and precision, and to adapt the value for $B$ to the input
value for $a/t$. For instance, for $a/t<1$, $B=20$ is sufficient
as $\mathcal{Z}(20)/\mathcal{Z}(0)$ is smaller than $10^{-14}$ for
moduli ratios up to 100.

\newpage
\part{Results}

In the following we will consider our results for the equivalent
modulus, ${\mathcal{E}}_{\rm eq}^*$, and the normalized
penetration, $\Delta$. The substrate to layer moduli mismatch
ratio ($E_0^*/E_1^*$) spans the range $10^{-2}$-$10^{2}$.

\section{Equivalent modulus}

\subsection{The equivalent homogeneous material}
For a given penetration $\delta$, let us define the ``equivalent
homogeneous material'' of a coated substrate as the homogeneous
material which would need the same applied load $P$ to get the
same penetration $\delta$ with the same indenter tip. Obviously,
this material would have a modulus equal to the ``equivalent
modulus'' of the coated substrate for the given penetration.

\subsection{Comparison with the Gao model}
We computed the equivalent modulus for flat-punch, sphere and cone
indentations for several moduli mismatch ratios ranging from
$10^{-2}$ to $10^{2}$. The comparison of the three sets of curves
shows that the same coated system being indented with different
indenter tips (flat-punch, cone or sphere) does not return the
same equivalent modulus for a given contact radius. However, the
qualitative evolution of the curves with the mismatch ratio
($E_0^*$/$E_1^*$) is the same, and  sphere and cone indentation
curves almost superimpose (Fig. 2).

Thus Fig. 3 only shows the evolution, for several mismatch ratios,
of the equivalent modulus with the relative contact radius for
cone indentations. Fig. 3(a) details the case of soft films on a
stiff substrate, whereas Fig. 3(b) represents the case of stiff
layers on a soft substrate. In all cases, the equivalent modulus
exhibits a transition from layer to substrate modulus as the
contact radius goes from zero to infinity.

One can notice that, whatever the indenter shape, all curves
almost have the same shape, regardless of the moduli mismatch
between the layer and the substrate. This shape indeed is very
much like that provided by Gao et al.$^6$. In particular, when the
moduli mismatch ratio converges to one, our curves converge to the
Gao model (Fig. 4). The agreement is good in the range 0.5-2. When
the mismatch is larger, the shape of the $E^*_{\rm eq}$ curves is
almost unchanged but the transition range, over which the system
response changes from one limit behavior to the other, appears to
shift from the position given by Gao. When indenting a soft layer
on a stiff substrate, the range over which the behavior of the
system is close to that of the film increases with the modulus
mismatch, whereas it conversely shrinks when indenting stiff
layers.

The analytical model introduced by Gao et al.$^6$ is based on a
perturbative analysis. Thanks to FE calculations, they showed that
their model is correct on the range 0.5-2. Thus, the agreement of
our results in the same range was to be expected.

As to the shift of the transition range for large modulus
mismatch, we note that in their analysis, Gao et al. used the
stress and displacement values calculated for an homogeneous
material. Thus, their analysis accounts for a homogeneous
distribution of the strain in the coated substrate. However, the
strain caused by indentation is a priori distributed
inhomogeneously in an inhomogeneous system. Indeed, as one of the
materials is softer than the other one, it tends to absorb a
greater part of the global strain. This means that the softer
material always dominates in the compound system response. Thus,
when indenting a soft film on a stiff substrate, most of the
strain being ``confined'' in the layer, the transition is shifted
towards greater relative contact radii. Conversely, when indenting
a stiff layer, it is the substrate which absorbs most of the
strain and the transition range is shifted to smaller $a/t$.
Naturally, this effect tends to increase with increasing modulus
mismatch.

As a result, the curve provided by the Gao model stands as a limit
between stiff layers and soft layers in our calculations (Fig.
4(b)). We are now in position to extend empirically the Gao
function on a wider range of moduli ratios.
\newpage

\subsection{Equivalent modulus : Empirical description}

As the $({E^*_{\rm eq}};a/t)$ curves resulting from our
calculations exhibit similar shapes, it appears reasonable to look
for a fit function that would expand the Gao function to a larger
range of moduli ratios. We propose the following function
\nolinebreak:
$${{E^*_{\rm eq}}}\left({\frac{a}{t}}\right)={{E^*_1}}+{\frac{({{E^*_0}}-{{E^*_1}})}{1+{({\frac{tx_0}{a}})}^{n}}}$$
where $x_0$ and $n$ are adjustable constants.

The parameter $x_0$ is the value of the $a/t$ ratio for which
${{E^*_{eq}}}=({E^*_0}+{E^*_1})/2$. At the same time it
corresponds to the change in curvature of the $({E^*_{\rm
eq}};a/t)$ curve plotted in semi-log coordinates.

Table I presents the results of our fit for cone contacts. The
parameter $x_0$ (which characterizes the position of the
transition range) increases with moduli mismatch ratio, as was
expected. Using our results we obtained the following relation on
a wide range of moduli mismatch :
$$\log(x_0)=-0.093+0.792\log({E_0^*}/{E_1^*})+0.05\left(\log({E_0^*}/{E_1^*})\right)^2$$

On the other hand, $n$ (which characterizes the width of the
transition range) does not change much on the $10^{-2}-10^2$
range.

\section{The real penetration}
The contact is described by three variables: the force, the
penetration and the radius of the contact zone, or contact radius.
For elastic systems, there exist two relations between these three
variables. For homogeneous half-spaces, these are the two
well-known relations due to Hertz in the case of a sphere. The
first relation links force and penetration. In the case of a
coated substrate, this relation leads to the introduction of the
effective elastic modulus which we have discussed so far. The
second relation links penetration and contact radius. It is as
simple as $\delta=\pi a/2 \tan\omega$ in the case of a cone. Does
this simple result remain unchanged when the substrate is
inhomogeneous?

For a coated substrate, we calculated the actual force and
penetration as a function of the contact radius. We defined the
normalized penetration $\Delta$ as the ratio of the actual
penetration, $\delta$, to the equivalent penetration $\delta_{\rm
eq}$ that would be obtained for the same given contact radius $a$,
with an homogeneous material (See Fig. 5). Thus, when $\Delta$ is
close to 1, the coated material behaves as if homogeneous. A
behaviour specific to coated substrates is evidenced when
$\Delta$ departs from  1.

\subsection{Evolution of the actual penetration during indentation}

$\Delta$ as a function of $a/t$ is plotted on Fig. 6 for cone and
sphere indentation. The case of flat-punch indentation is
irrelevant here as $a$ is a fixed parameter (Appendix A). Fig.
6(a) shows the evolution of $\Delta$ for the indentation of soft
layers, whereas Fig. 6(b) represents the indentation of hard
layers. In both graphs, the dashed and marked lines represent the
sphere indentations (right scale) while solid lines represent the
cone (left scale).

It can be noticed that, for these two indenter shapes, the
$(\Delta;a/t)$ curves are similar for a given moduli mismatch.
Indeed, one set of curves can be approximately re-scaled on the
other by $\Delta^{\rm sphere}-1=\kappa \left(\Delta^{\rm
cone}-1\right)$ where the geometrical factor $\kappa$ is around
1.37.

% As this curve is a plot of the $\delta - a$ relation, it is similar to the relation $h_c(a)=\delta-\varepsilon P/S$ in Oliver-Pharr method (where $P$ is the applied load, $S$ the contact stiffness and $h_c$ the contact depth). As this relation, in the case of homogeneous material, depends on the geometry of the indenter through $\varepsilon$ while keeping the same general shape, it appears reasonable that its extension in the case of a coated substrate behave the same way. Thus, the $\kappa$ factor should be interpreted in terms of a geometrical factor.

Fig. 6 shows that effects specific to coated materials appear for
an intermediate range of $a/t$ values. The width of this range
increases with the moduli mismatch. A superposition of Fig. 3 to
Fig. 6 provides a simple interpretation of this phenomenon.

Let us consider the case of the indentation of a soft film on a
stiff substrate (Fig. 6(a)). The equivalent penetration tallies
with the actual penetration for very small relative contact radii
because the substrate effect is then just a perturbation, and thus
can be neglected in first approximation. Similarly, when the
relative contact radius is very large, the system behaves as if
there were no film. In the intermediate range, for a given $a$,
$\Delta < 1$ means that the effective penetration is smaller than
the homogeneous equivalent penetration for the same given contact
radius $a$. Then, for a given $\delta$, the actual contact radius
is greater for the coated substrate than it is for the homogeneous
equivalent material. In fact, the applied strain is somehow
confined into the layer material, which elastically piles up
causing the contact radius $a$ to increase (and so the reduced
penetration $\Delta$ to decrease). When the stress applied on the
film becomes important enough, the substrate begins to distort
substantially, the elastic pile-up is less significant and
$\Delta$ increases, until the penetration is so important that the
film effect becomes negligible. Then the system turns back to
homogeneous substrate behavior. Conversely, for stiff films on
soft substrates, a larger penetration at identical contact radius
is obtained because of the compliance of the substrate
($\Delta>1$).

Finally, we conclude that, in the case of coated materials, even
if a contact radius dependent equivalent modulus can be defined,
the $\delta-a$ relation valid for homogeneous systems is no longer
systematically observed. This effect, previously described by
El-Sherbiney and Halling$^{15}$ apparently without much echo,
cannot be evidenced in flat punch calculations as $\delta$ is
independent from $a$. In fact, to mimic the definition of an
equivalent modulus, we would have to introduce a
penetration-dependent ``equivalent geometrical parameter'' equal
to the radius (resp. the semi-included angle) of the indenter tip
that would give the same $\delta-a$ relation if the system were
homogeneous.

A few qualitative considerations can be made on the evolution of
this parameter. First, the range of relative radii on which it
departs from the actual radius (resp. semi-included angle)
 corresponds to the transition range previously introduced for the
$(E_{\rm eq}^*;a/t)$ curves (See Fig. 3 and Fig. 6). Secondly, the
greater the moduli mismatch, the larger the maximum deviation of
the equivalent geometrical parameter from its actual value
becomes.

\subsection{Experimental methods, analytical models and actual penetration}

Let us recall two of the basic relations that are used in the
Oliver and Pharr depth-sensing indentation method$^1$ :
\begin{equation}
    {E}^*={\frac{\sqrt{\pi}S}{2\beta \sqrt{A}}}
\end{equation}
\begin{equation}
    h_c=\delta - \varepsilon {\frac{P}{S}}
\end{equation}
where $S$ is the contact stiffness, $A$ the contact area, $h_c$
the contact depth and $\beta$ and $\varepsilon$ parameters
depending on the shape of the indenter tip.

Let us comment on these equations: the discussion deals with the
cone, but the case of the sphere is similar. When indenting a real
system with a sharp indenter, the core problem to evaluate both
elastic and plastic properties is to separate elastic and plastic
contributions. Linear superposition shows that Eq. (24) holds for
an inhomogeneous substrate, provided ${E}^*$ is the flat punch
equivalent modulus for the contact radius $a$.

However, in the Oliver-Pharr method, the contact area is derived
from $h_c$ as calculated with Eq. (25). For an homogeneous
substrate, the validity of Eq. (25) stems from the fact that both
$\delta$ and $P/S$ are proportional to $a$. This is actually the
starting point for the calculation of the value of $\varepsilon$.
For a coated substrate, we have shown that $\delta$ is no longer
proportional to $a$. Similarly, $P/S$ also deviates from a linear
behaviour. Eq. (25) is therefore likely to break down, which we
checked in a few sample cases.

Thus, for significant modulus mismatch, the contact radius $a$
calculated from Eq. (25) may be in error by up to 25\% for an
elastic modulus mismatch of 100 (See Fig. 6(a)). This error will
propagate to the effective modulus derived from Eq. (24) and to
the layer modulus inferred from the effective modulus.

\section{Conclusion}

We introduced an analytically exact approach based on the Green
function for coated materials calculated by Li and Chou$^{10}$.
Using a basis of functions well-suited for the mixed boundary
conditions, we obtained an integral relation that was easily
solved numerically. Based on this relation, we set up an algorithm
which provides, given the contact radius, applied force and
penetration, for arbitrary moduli ratios and arbitrary
axisymmetric indenter shape.

For moduli mismatch ratios close to 1, our results showed good
agreement with the Gao model. On a wider range of moduli ratios,
we have shown that, for soft layers (resp. stiff layers), the
transition from layer to substrate-dominated behavior shifts to
larger contact radii (resp. smaller contact radii) with increasing
moduli mismatch. We rationalized this results in terms of
inhomogeneous strain distribution in the coated material. Based
on our computations, we provide an empirical function to describe
the (${E^*_{\rm eq}}$; $a/t$) curves on a large range of modulus
mismatch ratios.

In addition, we have evidenced that, for indention of a coated
material with a non-flat indenter, the penetration to contact
radius relation significantly deviates from what it is for a
homogeneous substrate. This result casts some doubts on the
validity of the Oliver and Pharr equations in coated systems
mechanical evaluation. On-going work investigates the application
of the present method to an Oliver-Pharr-like data treatment for
coated material.

\section*{ACKNOWLEDGMENTS}

The authors would like to acknowledge St\'ephane Roux for his help
and advice, particularly on the mathematical and numerical aspects
of the algorithm.

%\bibliographystyle{prsty}
%\bibliography{igorbib}
\newpage

\section*{REFERENCES}

\begin{enumerate}

\item W. C. Oliver and G. M. Pharr, J. Mater. Res. 7 ,1564 (1992).

\item H. Hertz, J. Reine und angewandte Mathematik 92, 156 (1882).

\item I. N. Sneddon, {\em Fourier Transforms} (McGraw-Hill Book
Company, Inc., New York, 1951).

\item M. F. Doerner and W. D. Nix, J. Mater. Res 1, 601 (1986).

\item J. Men$\breve{c}$ik, D. Munz, E. Quandt, E. R. Weppelmann
and M. V. Swain, J. Mater. Res 12, 2475 (1997).

\item H. J. Gao, C. H. Chiu and J. Lee, Int. J. Solids Structures
29, 2471 (1992).

\item E. H. Yoffe, Ph. Mag. Let 77, 69 (1998).

\item N. Schwarzer, Journal of Tribology 122, 672 (2000).

\item V. I. Fabrikant, {\em Application of Potential Theory in
Mechanics : A Selection of New Results} (Kluwer Academic
Publishers, The Netherlands,  1989).

\item J. Li and T. W. Chou, Int. J. Solids Structures 34, 4463
(1997).

\item I. N. Sneddon, Int. J. Engng. Sci. 3, 47 (1965).

\item A. Huguet and E. Barthel, J. Adhesion 74, 143 (2000).

\item G. Haiat and E. Barthel, Langmuir 18, 9362 (2002).

\item J. C. Hay, A. Bolshakov and G. M. Pharr, J. Mater. Res. 14,
2296 (1999).

\item M. El-Sherbiney and J. Halling, Wear 40, 325 (1976).

\end{enumerate}
\newpage

\section*{CAPTIONS}
{\underline{Fig.1 :}} Schematic representation of the indentation
of a coated elastic half-space.\\

{\underline{Fig.2 :}} Evolution of the equivalent modulus ${E_{\rm
eq}}^*$ with $a/t$ for flat-punch (dashed), cone (plain) and
sphere (markers) indentation for
${E_{0}}^*/{E_{1}}^*=0.1$($\times$) and
${E_{0}}^*/{E_{1}}^*=10$($+$). All curves are
different but cone and sphere indentation almost coincide.\\

{\underline{Fig.3 :}} Evolution of the equivalent modulus with the
moduli mismatch ratio in the case of the indentation by a cone of
soft layers (a)($E_0^*/E_1^*=2$ ($+$); 5 ($\blacksquare$); 10
($\blacktriangle$) ; 100 ($\times$)) and stiff layers
(b)($E_0^*/E_1^*=0.5$ ($+$); 0.2 ($\blacksquare$); 0.1
($\blacktriangle$) ; 0.01 ($\times$)). The bold dashed line is the Gao function$^6$.\\

{\underline{Fig.4 :}} Evolution of the equivalent modulus with the
contact radius in the case of moduli mismatch($E_0^*/E_1^*$) of
0.5 ($\blacktriangle$), 0.9 ($\times$), 1.1 ($+$) and 2
($\blacksquare$) for cone indentations (a) and flat punch
indentations (b). Our results converge to the Gao model (bold
dashed line) when the moduli mismatch tends to 1. Moreover, the
transition range for stiff layers (resp. soft layers) is shifted
towards smaller (resp. larger) values of $a/t$,
in comparison to the transition range given by the Gao model.\\

{\underline{Fig.5 :}} Schematic representation of the indentation
of a coated substrate and its equivalent homogeneous material in
the case of a soft layer on a stiff substrate ($\Delta<1$). For
the same given contact radius,
two different penetrations are obtained.\\

{\underline{Fig.6 :}} Evolution of the normalized penetration
$\Delta$ in our calculation for soft layers (a)($E_0^*/E_1^*=2$
($+$); 5 ($\blacksquare$); 10 ($\blacktriangle$) ; 100 ($\times$))
and stiff layers (b)($E_0^*/E_1^*=2$ ($+$); 5 ($\blacksquare$); 10
($\blacktriangle$) ; 100 ($\times$)) for sphere (dashed and
marked)
and cone (solid line) indentations. The same system indented with two different indenters returns curves which can be approximately scaled onto each other by a scaling factor of 1.37.\\

\newpage
\section*{TABLES}
\begin{center}
\begin{tabular}{|c|c|c|c|c|c|}
  \hline
  % after \\: \hline or \cline{col1-col2} \cline{col3-col4} ...
  $E_0^*/E_1^*$ & $x_0$ & $n$ & $E_0^*/E_1^*$ & $x_0$ & $n$ \\
  \hline
  100 & 52.95 & 1.06  & 0.67 & 0.60 & 1.27  \\
  50 & 25.17 & 1.09  & 0.5 & 0.48 & 1.27  \\
  25 & 12.55 & 1.13  & 0.2 & 0.24 & 1.29  \\
  10 & 5.36 & 1.18 & 0.1 & 0.14 & 1.31  \\
  5 & 2.96 & 1.22  & 0.04 & 0.082 & 1.32  \\
  2 & 1.41 & 1.26  & 0.02 & 0.052 & 1.31  \\
  1.5 & 1.13 & 1.27  & 0.01 & 0.032 & 1.27  \\
  \hline
\end{tabular}
\end{center}
\,

\,

 {\underline{Table I :}} Values obtained for $x_0$ and $n$ while fitting the computed equivalent modulus curves for cone indentation (see text).

\newpage
\appendix
\section{Expressions for other commonly-used indenter shapes}

\subsection{Flat punch indenter}
Let us consider first the case of the flat punch indenter, which
is the simplest. As the displacement is constant, using (10) we
get :

\be \label{b0}\forall\, s \leq a,\, \theta(s)=\delta \ee

Combining Eq. (13) and Eq. (\ref{b0}), we get :

\begin{equation}\label{b1}
\forall\, s \leq a,\, \delta={\frac{2}{\pi}}\inta g(r) \left(\intt
dk\, \mathcal{C} (kt) \cos(kr) \cos (ks)\right)dr
\end{equation}

With a normalization similar to that introduced for the cone, we
get the following normalized quantities :

\be\label{b2}
\rho\equiv{\frac{r}{a}}\,;\,\varsigma\equiv{\frac{s}{a}}\,;\,\tau\equiv{\frac{t}{a}}\,;\,\eta\equiv
ka \ee

\be\label{b3}
\mathcal{Z}(x)\equiv{\frac{E_1^*}{2}}\mathcal{C}(x)-1 \ee

\be\label{b4} G(\rho)\equiv{\frac{2 \, g(\rho)}{\delta\, E_1^*}}
\ee

Introducing these quantities in Eq. (\ref{b1}), we get :

\begin{equation}\label{b5}%24
\forall \,\varsigma \leq 1,\,
1=G(\varsigma)+{\frac{2}{\pi}}{{\int_{0}}^{1}}G(\rho) \left(\intt
d\eta\, \mathcal{Z} (\eta \tau) \cos(\eta \rho) \cos (\eta
\varsigma)\right) d\rho
\end{equation}

Through this normalization, we can get the expression of a
normalized applied load and a normalized equivalent elastic
modulus, thanks to the expression of $G(\rho)$.

\begin{equation}\label{b6}%22
\Pi=4{{\int_{0}}^{1}}d\rho\, G(\rho)
\end{equation}

\begin{equation}\label{b7}%23
{\mathcal{E}}_{\rm eq}^*={\frac{E_{\rm
eq}^*}{E_1^*}}={\frac{\Pi}{4}}
\end{equation}
\\
\subsection{Sphere indentation}
In the case of a sphere indentation, we have
$u(r)=\delta-r^2/(2R)$, $R$ being the radius of the sphere. We
thus can write, using Eq. (10), that under the contact:

\begin{equation}\label{b8}%13
\forall\, s \leq a, \,  \theta(s)=\delta-{\frac{s^2}{R}}
\end{equation}
\newpage
Similarly to the case of the cone, we introduce the following
normalized quantities\nolinebreak :

\be\label{b9}
\rho\equiv{\frac{r}{a}}\,;\,\varsigma\equiv{\frac{s}{a}}\,;\,\tau\equiv{\frac{t}{a}}\,;\,\eta\equiv ka\\
\ee \be\label{b10}
\Delta \equiv{\frac{\delta R}{a^2}}\\
\ee

\be\label{b11}
\mathcal{Z}(x)\equiv{\frac{E_1^*}{2}}\mathcal{C}(x)-1 \ee

\be\label{b12} G(r)\equiv{\frac{2R\,g(r)}{a^2\,E_1^*}} \ee

We then get :

\begin{equation}\label{b13}
\forall\, \varsigma \leq 1,\,
\Delta-{\varsigma}^2=G(\varsigma)+{\frac{2}{\pi}}{{\int_{0}}^{1}}
G(\rho) \left(\intt d\eta\, \mathcal{Z} (\eta \tau) \cos(\eta
\rho) \cos (\eta \varsigma)\right)d\rho
\end{equation}

\begin{equation}\label{b14}
\Pi={\frac{2PR}{a^3\,E_1^*}}=4{{\int_{0}}^{1}}d\rho\, G(\rho)
\end{equation}

\begin{equation}\label{b15}%17
{\mathcal{E}}_{\rm eq}^*={\frac{E_{\rm
eq}^*}{E_1^*}}={\frac{3}{8}}{\frac{\Pi}{{\Delta}^{{3/2}}}}
\end{equation}

\newpage

\begin{figure}
% Requires \usepackage{graphicx}
\includegraphics[width=3.25in]{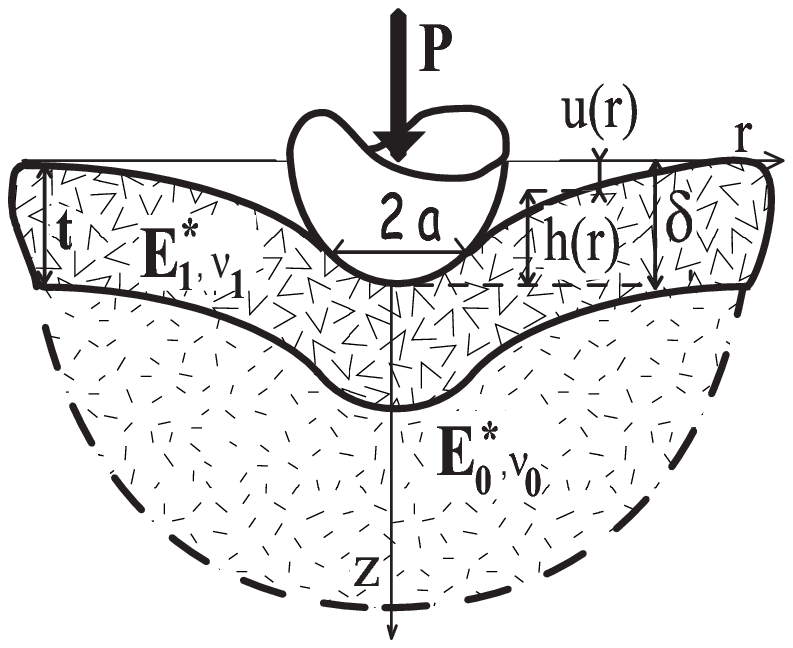}
\caption{}\label{}
\end{figure}

\newpage

\begin{figure}
% Requires \usepackage{graphicx}
\includegraphics[width=3.25in]{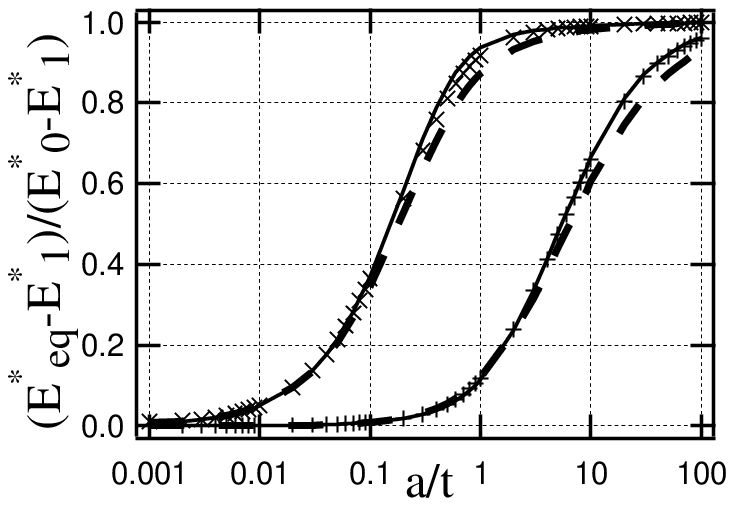}
\caption{}\label{}
\end{figure}

\newpage
\begin{figure}
% Requires \usepackage{graphicx}
\includegraphics[width=3.25in, height=2.25in]{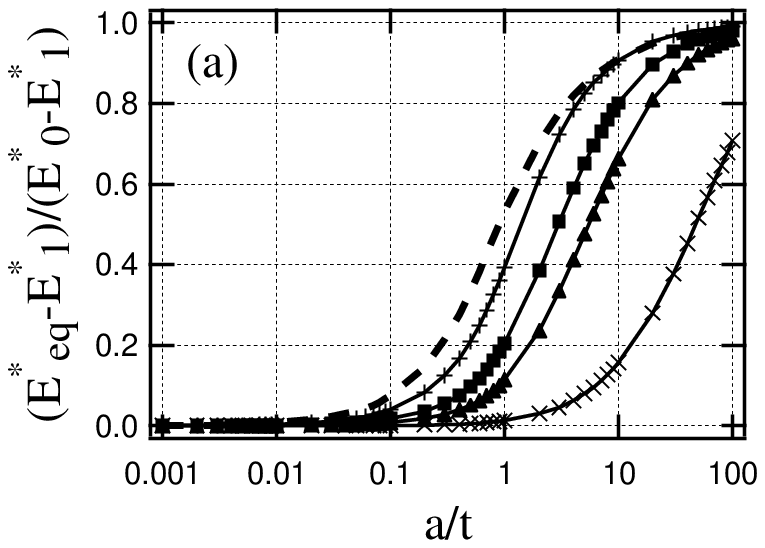}
\\
\includegraphics[width=3.25in, height=2.25in]{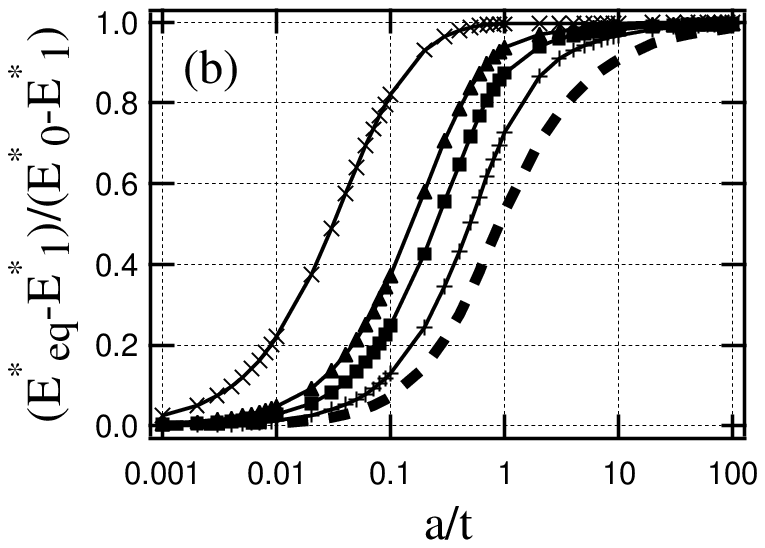}
\caption{}\label{}
\end{figure}
\newpage

\begin{figure}
% Requires \usepackage{graphicx}
\includegraphics[width=3.25in, height=2.25in]{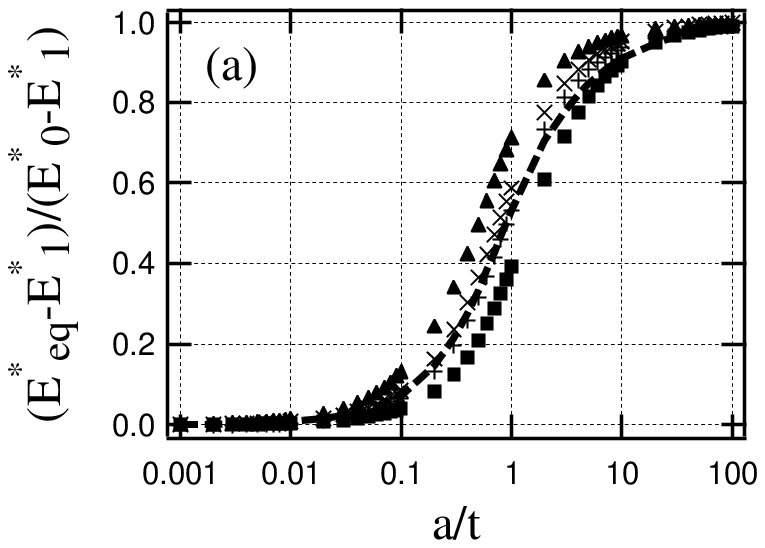}
\\
\includegraphics[width=3.25in, height=2.25in]{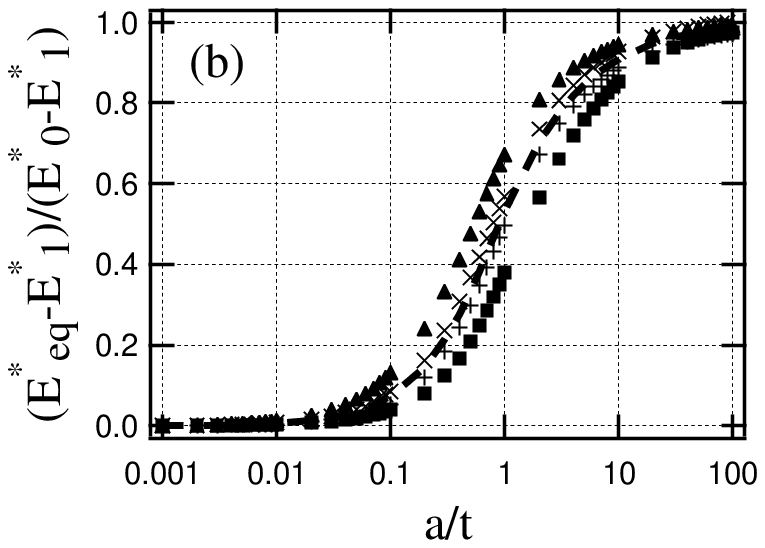}
\caption{}\label{}
\end{figure}
\newpage

\begin{figure}
% Requires \usepackage{graphicx}
\includegraphics[width=3.25in]{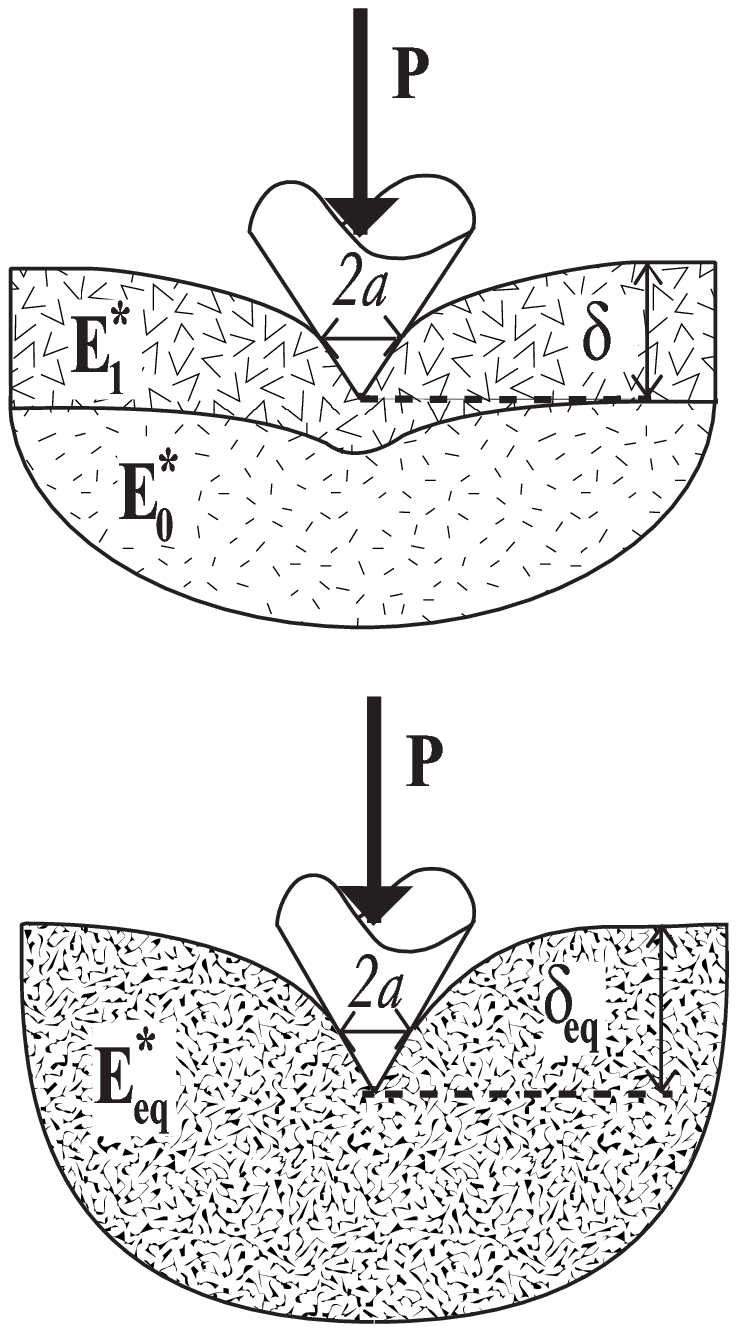}
\caption{}\label{}
\end{figure}
\newpage

\begin{figure}
% Requires \usepackage{graphicx}
\includegraphics[width=3.25in, height=2.25in]{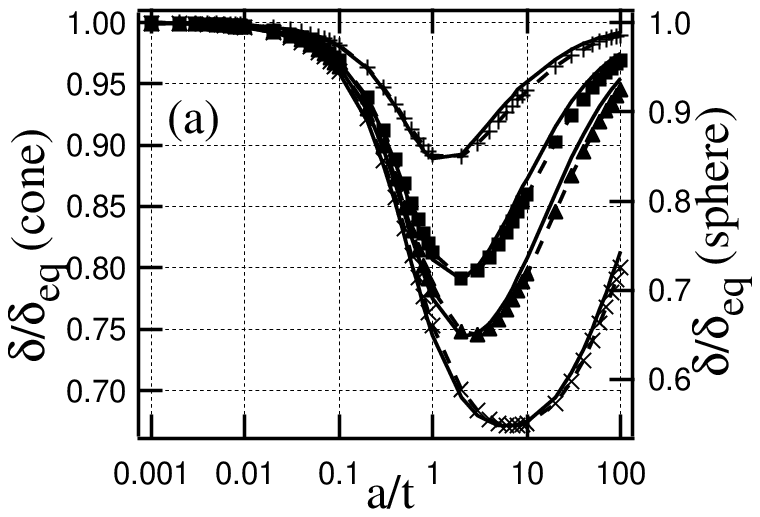}
\\
\includegraphics[width=3.25in, height=2.25in]{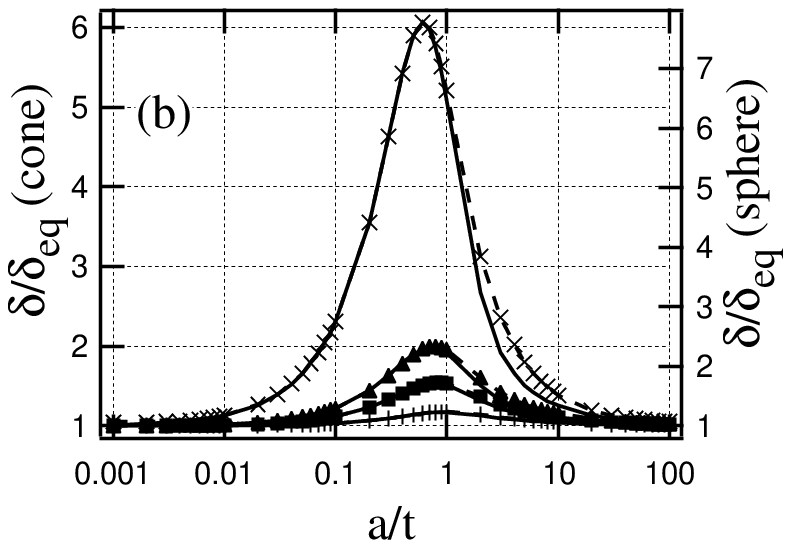}
\caption{}\label{}
\end{figure}

\end{document}